\documentclass[manuscript]{aastex}
\usepackage{pdfpages}
\usepackage{graphicx}
\usepackage{pdflscape}
\usepackage{afterpage}

\newcommand{\ce}{E(B$-$V)}

\shorttitle{Broad Diffuse Bands}
\shortauthors{Galazutdinov et al.}

\begin{document}

\title{Survey of very broad Diffuse Interstellar Bands}

\author{G. Galazutdinov}
\affil{Instituto de Astronomia, Universidad Catolica del Norte, Av. Angamos 0610, Antofagasta, Chile\\
Pulkovo Observatory, Pulkovskoe Shosse 65, Saint-Petersburg 196140, Russia}
\email{runizag@gmail.com}

\author{A. Bondar}
\affil{ICAMER Observatory, NAS of Ukraine, Acad. Zabolotnoho 27, 03143, Kyiv, Ukraine}
\email{arctur.ab@gmail.com}

\author{Byeong-Cheol Lee}
\affil{Korea Astronomy and Space Science Institute, 776, Daedeokdae-ro, Yuseong-gu, Daejeon, Korea 34055}
\email{bclee@kasi.re.kr}

\author{R. Hakalla, W. Szajna and J. Kre{\l}owski}
\affil{Institute of Physics, University of Rzesz{\'o}w, Pigonia 1 Street, 35-310, Rzesz{\'o}w, Poland}
\email{hakalla@ur.edu.pl; szajna@ur.edu.pl; jacek@umk.pl}

\begin{abstract}
This paper considers a very special set of a few interstellar features --- broad diffuse interstellar bands (DIBs) at 4430, 4882, 5450, 5779 and 6175 \AA\AA. The set is small, and
measurements of equivalent widths of these DIBs are challenging because of severe stellar, interstellar, and sometimes, also telluric contaminations inside their broad profiles.
Nevertheless, we demonstrate that they do correlate pretty tightly (DIBs 4882 and 5450 to a lesser extent though) with other narrower diffuse bands, as well as with the color excess E(B$-$V).
The studied broad  DIBs correlate  well with both interstellar molecule CH and interstellar K{\sc i},  i.e. it is hardly possible  to verify whether the environments, facilitating the formation
of very broad DIB carriers, are dominated by either molecular or atomic gas  as both these species likely occupy the same volume.
\end{abstract}

\keywords{ISM: clouds  --- ISM: lines and bands --- ISM: molecules ---
ISM: individual objects (very broad DIBs)}

\section{Introduction}

The diffuse interstellar bands (DIBs) have been known since nearly 100 yr \citep{Heg22} but still
remain unidentified. For a recent review see \cite{Kre18}.  The recently promoted identification of a few near infrared DIBs
as being carried by C$_{60}^+$ (\cite{CLC19} and references therein) does not take into account the fact that the two strongest DIBs
do not show the laboratory predicted strength ratio and thus those of weaker bands are very uncertain (\cite{GaS17}, \cite{GK17}).  DIB profiles are
always broader than those of interstellar atomic lines or those of simple interstellar radicals (CH, CN, CH$^+$).  The most recent
survey of DIBs (Fan et al., 2019) reports the presence of 556 features in the optical range. However, several DIBs have been discovered
also in the infrared part of spectrum, see e.g. \cite{GaK17} and references therein.
In turn, diffuse bands are tentatively divided into two or three non-equal groups: the first group is the most numerous one of so-called narrow
diffuse bands with a typical  full width at the half maximum (FWHM) close to $\sim$1 \AA\, the second contains a few features with FWHM$\sim$4 \AA, and the third
one is a small group of broad DIBs with FWHM $\geq$ 10~\AA. In the current research we focus on it. Let us emphasize
that the difference  between broad and narrow diffuse bands is of physical origin. Perhaps, they are formed by different kinds of molecules.
Indeed, the broad DIBs are  well seen in targets being in the environment characterized by  strong UV flux, e.g. in $\sigma$ type objects, where for
so-called narrow DIBs, the features that seem to be related to absorption features of simple
interstellar radicals (CH, CN, CH$^+$) are very weak (see \cite{Kre18} and references therein).

The first very broad DIB, centered at $\sim$4430~\AA\, was identified as such more than 80 yr ago
(\cite{BB38}; \cite{MH38}). The band, as broad as about 40~\AA, FWHM $\sim$ 20~\AA\,
was found as stationary in a spectroscopic binary. It was reasonably easy to demonstrate that the
strength of this feature correlates with a color excess but precise measurements were difficult
since the very broad profile was naturally contaminated (especially in B-type stars) with numerous
stellar lines of various intensity \citep{Her66}. This is why the first determined correlations between 4430 and
color excess or intensities of other interstellar features were found to be reasonably poor
\citep{GA50}.

The discovered later, 6170 DIB, is the only very broad DIB that is not severely contaminated with
stellar lines \citep{Ru70}. The latter publication reported the 4430/6175 strength ratio
to  be related to the galactic longitude and to the total-to-selective extinction ratio. The conclusion was,
however, based on a rather small sample of reddened stars. We found that the 6170 diffuse band is often blended with
another broad DIB marked as 6177 \AA\ one. Also, this double-DIB is often blended with many narrow diffuse bands
(Galazutdinov et al., 2000, Hobbs et al., 2008), challenging the measurements.
Here we measured  the features of 6170 and 6177 together and marked the resulting one as 6175 DIB.

Diffuse bands' intensities, especially of 4430, have been typically
related to the continuous extinction curve \citep{Her67}.  Until the end of the 1960s the 4430 DIB was the most
frequently investigated diffuse feature. The band is the first DIB observed outside our Milky Way system
\citep{HNM80}. Until the beginning of the 1980s the
hypothesis that the DIBs are carried by dust grains was taken very seriously. \cite{Her67}
compared the problem of DIB identification to those of nebulium and coronium, saying
that its solution may be of similar importance.

A statistical investigation of 65 sources revealed a rather tight correlation between the color excess and the
4430 DIB intensity \citep{Gam75}. The relation suffers, however, a rather serious scatter among heavily reddened targets.
Soon after the extra-atmospheric observations have been launched, \citet{Dan80} suggested the correlation between
the 4430 DIB and the 2200 bump of the extinction curve. \citet{ISNT86} analyzed a sample of 482
reddened stars. Their 4430 strengths do correlate with \ce\ but the scatter is very substantial.

A real breakthrough in the DIB investigations was the publication of \cite{Her75} which, due to the enhanced
signal-to-noise (S/N) ratio, raised the number of known DIBs from 9 to 39 features plus a few suspects. Herbig 
demonstrated quite narrow DIBs (like 6196 or 6379), as well as pretty broad (like 5780 or 6284) and very broad 
ones (like 4882 or 6175). All these features' intensities showed a rather tight
correlations with the \ce. The Herbig's investigation was based on averaged spectra with S/N$\sim$100.
A statistically valuable sample of the 4430 central depth in low-resolution, high-S/N spectra
\citep{KWGH87} suggested the local differences in relative strengths of the very broad feature. The paper emphasized
the problem of the profile stellar contaminations. Later the very broad DIBs were investigated rather seldomly.

The recent paper of \citep{SYH18} revitalized the idea of investigating the
broadest diffuse bands. They studied  21 sightlines and demonstrated profiles of a few broad DIBs,
but only the relation of 6175 to \ce. Moreover, they analyzed
correlations between the broad DIBs and narrower ones. In their opinion
the 4963 DIB originates in rather molecular gas while 5780, 5797, 6284,
and 6613~\AA\ DIBs primarily trace atomic gas. Our idea is to check the
above-mentioned suggestions using a larger sample of much higher
resolution spectra.

\section{Observational data}

Our observations have been collected using several high-resolution, fiber-fed
echelle spectrographs and are as follows.
\begin{itemize}

\item The Fiber-fed Extended Range Optical Spectrograph (FEROS), being fed with the 2.2m ESO L
aSilla telescope \citep{Kauf99} allows one to record in a single exposure the spectral range from 3600 to
9200~\AA\, divided into 39 echelle orders. The resolution of the  FEROS
spectra is R=48,000. FEROS spectral orders cover pretty broad wavelength ranges, which
make the spectrograph a very useful tool to
check the spectral types and luminosity classes of the observed
targets. The measurements are marked as "F" in Table 1.
\item The Echelle SpectroPolarimetric Device for the Observation of Stars (ESPaDOnS)
 spectrograph is the bench-mounted high-resolution echelle
spectrograph/spec\-tro\-pola\-ri\-meter) attached to the 3.58~m
Canada-France-Hawaii telescope at Maunakea (Hawaii, US). It is
designed to obtain a complete optical spectrum in the range from
3700 to 10,050~~\AA. For details see {\it https://www.cfht.hawaii.edu/Instruments/Spectroscopy/ESPaDOnS/}.
The whole spectrum is divided into 40 echelle
orders. The resolving power is about 68,000. The spectra from
ESPaDOnS were obtained during the runs 05Ao5 (in 2010, with PI:
B.~Foing) and 15AD83 (in 2015, with PI: G.~Walker). The measurements are marked as "E" in Table 1.
\item The High Accuracy Radial velocity Planet Searcher (HARPS)
 spectrograph \citep{May03}, fed by the 3.6m ESO LaSilla
telescope has a resolving power
 R=115,000 and broad spectral range which allows to cover all spectral
lines used for classification of stars: He{\sc i}, He{\sc ii}
and Mg{\sc ii} as well as the interstellar Ca{\sc ii} lines.
HARPS spectra cover also a vast majority of Diffuse Bands and
interstellar molecular features. The measurements are marked as "H" in Table 1.
\item
BOES (Bohyunsan Echelle Spectrograph) of the Korean National
Observatory \citep{kimetal2007} is installed at the 1.8m telescope
of the Bohyunsan Observatory in Korea. The spectrograph has three
observational modes allowing resolving powers of 30,000, 45,000 and
90,000. In any mode, the spectrograph covers the whole spectral range from
$\sim$3500 to $\sim$10,000~\AA, divided into 75 -- 76 spectral
orders. The measurements are marked as "B" in Table 1.
\end{itemize}

\begin{figure}
\plotone{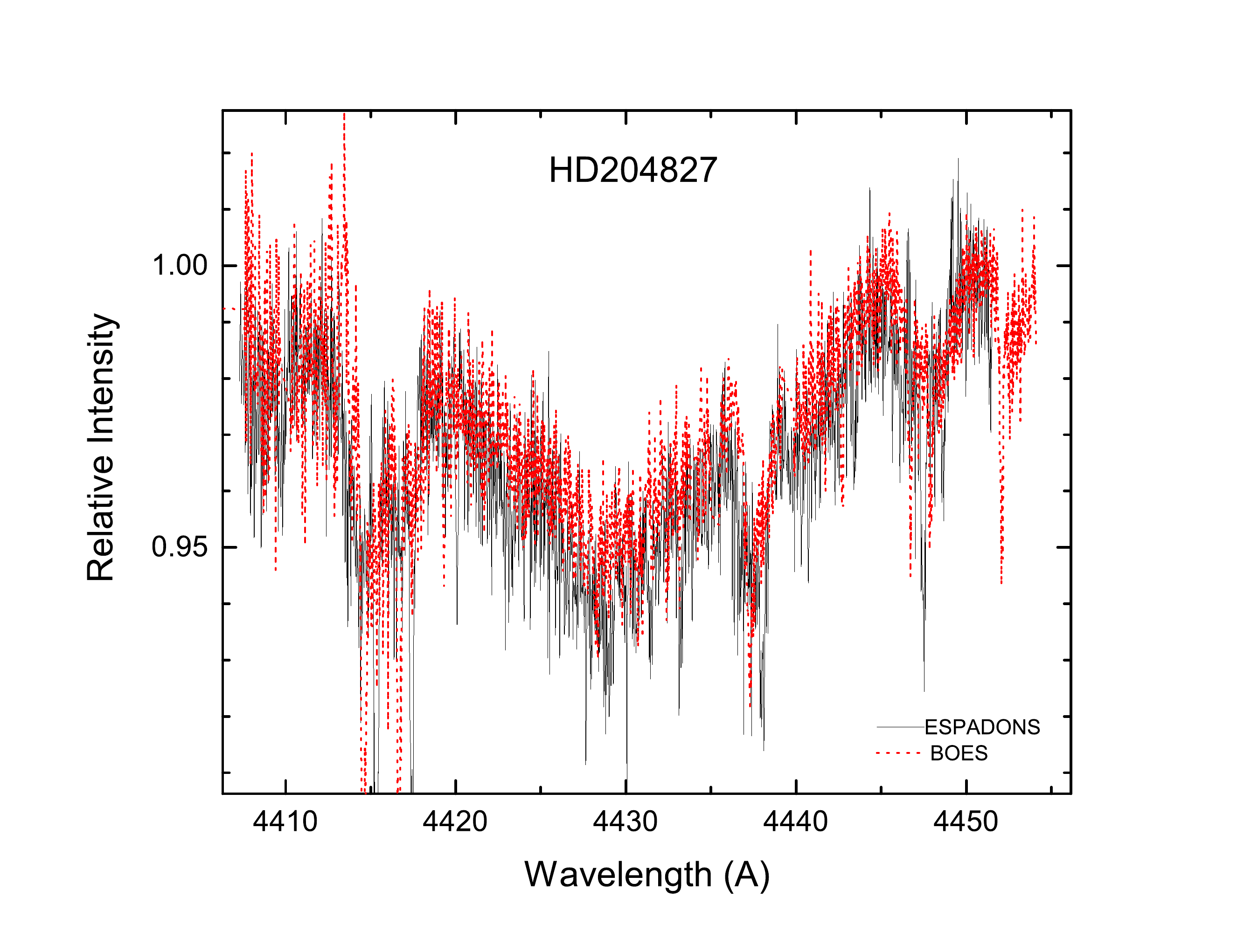}
\caption{Comparison of the 4430 DIB profiles in the spectra from fiber--fed spectrographs
ESPaDOnS and BOES. Note their identity inside the apparent noise.}
\label{compar}
\end{figure}

It is worth mentioning that all of the above spectrographs are fiber--fed, in most of cases providing an almost excellent
flat-fielding procedure. The resulting flat-fielded spectra normally are nearly flat, i.e. the continuum is
almost a straight line, making the normalization procedure much more evident than for those of slit instruments in most of cases.
This is a very important advantage of fiber-fed spectrographs.
Spectra from high-resolution slit spectrographs are often hardly useful for reliable measurements of
broad features due to the complex shape (rather crowded) of spectral orders after the flat-field normalization.
Also, we used the manual fit procedure to restore the broad DIB profiles, which allows for the elimination of stellar contamination.
To confirm the above we present the plot (Fig.~\ref{compar}) comparing the HD204827 spectrum
acquired with two fiber--fed echelle spectrographs: BOES and ESPaDOnS.
In Fig.4 we show how we measured DIB5780 blended with broad DIB5779: the profile of the broad feature was accepted as a kind of pseudo-continuum.
All other blended features were measured in the same manner.

 Some objects were observed more than once. For the analysis we have selected the spectra with higher S/N ratio and/or with more
evident position of the continuum.
 The complete set of measured profiles of broad diffuse bands is given in the Appendix.

All spectral images (except the ESPaDOnS data) were processed and measured in standard way using
our own
DECH\footnote[1]{available upon request} code.
The spectral resolutions, provided by the above mentioned instruments,
are not identical but all are high enough to precisely measure the strengths of the
investigated spectral features of atomic and molecular species and, especially, of the broad DIBs.

\afterpage{%
\clearpage
\begin{landscape}
\begin{deluxetable}{llc c c c c c}
\tablewidth{290pt}
\tabletypesize{\footnotesize}
\tablewidth{0pt}
	\tablecaption{Equivalent widths (m\AA) of the confirmed broad DIBs,
visible in the spectra of the selected targets, the signal-to-noise ratio  (S/N) and FWHM in \AA.
 The selected stars cover a wide range of spectral types (Sp/L~= O3I~-- B8Ib) and reddening (\ce~= 0.28~-- 2.01).
	\label{cdibs}}
	\tablehead{
	\colhead{Star}
        & \colhead{Sp/L}     &
                	\colhead{\ce}
                         & \colhead{4428.3 S/N FWHM} &
	                                                  \colhead{4882.0 S/N FWHM} &
                                                                      \colhead{5450.3 S/N FWHM} &
	                                                                                               \colhead{5779.1 S/N FWHM} &
                                                                                                                              \colhead{6175 S/N FWHM}
}
\startdata
BD+404220E&  O6V     &  1.96  & 2700$\pm$1300  50  18 & 4120$\pm$940  90  31 &  560$\pm$455 110  17& 1143$\pm$300 220 16& 1988$\pm$700 130 30 \\
BD-134923F  &  O5.5  &  1.10  & 1760$\pm$490  135  17 & 1310$\pm$450 213  35 &  375$\pm$130 275  14&  718$\pm$170 290 14& 1222$\pm$247 290 30 \\
BD-134927F  &  O7II  &  0.95  & 1540$\pm$362  135  16 & 1545$\pm$575 195  35 &  407$\pm$185 210  14&  870$\pm$270 250 19&  993$\pm$340 176 26 \\
BD-134928F  &  O9.5V &  0.77  & 1201$\pm$460  110  17 &                      &  395$\pm$200 200  13&  775$\pm$428 160 16&  690$\pm$337 170 27 \\
BD-134929F  &  O9V   &  0.90  & 1500$\pm$421  135  17 &                      &  356$\pm$240 155  14&  660$\pm$250 230 15&  968$\pm$317 185 28 \\
CygOB\_7B &  O3I     &  1.74  & 2340$\pm$1700  35  17 &                      &  435$\pm$280 120  14& 1366$\pm$390 130 19& 1758$\pm$459 170 19 \\
CygOB\_8B &  O6I     &  1.58  & 2160$\pm$510   90  16 & 1100$\pm$430 180  20 &  533$\pm$158 270  15&  905$\pm$230 265 14& 1352$\pm$240 300 21 \\
CygOB\_11B&  O5.5I   &  1.78  & 2900$\pm$900   45  18 & 1940$\pm$630 110  25 &  456$\pm$200 170  14& 1040$\pm$535 108 13& 1982$\pm$369 210 22 \\
CygOB\_12E&  B4Ia    &  2.01  &   \nodata             &                      &                     & 1150$\pm$890  65 14& 2060$\pm$1147 67 29 \\
Hersch36F &  O7.5V   &  0.90  & 1620$\pm$350  190  16 &                      &  388$\pm$74  470  14& 1075$\pm$364 250 18&  392$\pm$120 405 20 \\
 15785B   &  B1Iab   &  0.65  & 1400$\pm$240  250  16 &  640$\pm$220 400  32 &  323$\pm$99  350  13&  431$\pm$160 370 15&  480$\pm$180 320 24 \\
 24912H   &  O7.5III &  0.31  &  624$\pm$210  220  16 &                      &   83$\pm$78  370  14&  215$\pm$190 280 15&  457$\pm$230 290 13 \\
 34078B   &  O9.5V   &  0.49  &  760$\pm$170  310  19 &                      &                     &  129$\pm$62  800 13&  315$\pm$226 280 24 \\
 73882F   &  O8.5IV  &  0.68  &  840$\pm$270  280  15 &                      &   79$\pm$41  680  10&  170$\pm$120 450 15&  460$\pm$110 590 22 \\
 76341F   &  O9.2IV  &  0.51  &  680$\pm$350  180  19 &                      &  181$\pm$104 335  11&  130$\pm$120 350 12&  334$\pm$143 360 24 \\
 78344F   &  O9.5Ib  &  1.32  & 1906$\pm$232  270  16 & 2580$\pm$375 310  33 &  324$\pm$111 325  13&  640$\pm$137 425 15&  987$\pm$161 390 21 \\
 80077F   &  B2Ia    &  1.45  & 3100$\pm$1500  40  19 & 2890$\pm$1200 80  32 &  400$\pm$121 280  12& 1243$\pm$307 270 17& 1230$\pm$330 320 26 \\
147165H   &  B1III   &  0.34  &  714$\pm$236  220  16 &                      &                     &  227$\pm$67  810 12&                     \\
147888F   &  B3V     &  0.49  & 1000$\pm$150  510  18 &                      &  120$\pm$50  515  12&  514$\pm$124 550 17&                     \\
147889H   &  B2V     &  1.04  & 1490$\pm$281  280  18 &                      &  133$\pm$45  560  13&  652$\pm$166 425 17&                     \\
148379F   &  B2Iab   &  0.74  & 1665$\pm$150  524  16 & 1600$\pm$170  645 34 &  290$\pm$81  670  14&  450$\pm$108 570 13&  370$\pm$90  600 19 \\
148937F   &  O6I     &  0.65  & 1218$\pm$380  173  17 &                      &  174$\pm$152 240  10&  630$\pm$308 245 15&  809$\pm$267 300 26 \\
149038F   &  O9.7Iab &  0.28  & 1050$\pm$160  535  15 &                      &  127$\pm$75  550  13&  396$\pm$90  940 17&  406$\pm$136 495 19 \\
149404F   &  O8.5Iab &  0.60  & 1336$\pm$87   800  16 & 1140$\pm$140  720 28 &  304$\pm$109 515  15&  592$\pm$93  940 15&  746$\pm$130 500 22 \\
149757H   &  O9.5V   &  0.29  &  120$\pm$30   800  9  &                      &                     &                    &  128$\pm$56  600 14 \\
152233F   &  O6II    &  0.42  &  960$\pm$300  240  17 &                      &  174$\pm$102 350  13&  407$\pm$296 290 18&  445$\pm$90  560 10  \\
152235F   &  B0.5Ia  &  0.74  & 1300$\pm$100  650  14 & 1025$\pm$255 460  34 &  180$\pm$122 380  13&  457$\pm$134 480 15&  345$\pm$80  570 14 \\
152249F   &  O9Iab   &  0.44  & 1000$\pm$190  400  14 &  535$\pm$176 450  25 &  200$\pm$120 422  14&  280$\pm$102 470 11&  485$\pm$148 420 17 \\
154368F   &  O9.5Iab &  0.73  &  850$\pm$130  540  17 &  220$\pm$100 560  20 &  210$\pm$56  680  15&  354$\pm$67 1035 16&  457$\pm$200 300 18 \\
154445F   &  B1V     &  0.35  &  710$\pm$135  550  14 &                      &                     &  278$\pm$121 435 16&  110$\pm$90  470 14 \\
157038F   &  B4Ia    &  0.83  & 1550$\pm$100  690  17 & 1530$\pm$190 560  33 &  227$\pm$92  450  14&  413$\pm$182 330 12&  613$\pm$140 460 19 \\
163800H   &  O7.5III &  0.56  &  700$\pm$100  480  14 &                      &                     &  325$\pm$124 470 15&  470$\pm$97  565 12 \\
166734F   &  O7.5I   &  1.34  & 2560$\pm$230  320  17 & 2105$\pm$315 350  33 &  855$\pm$110 490  15& 1030$\pm$240 360 17& 1276$\pm$245 360 26 \\
168112F   &  O5.5    &  1.00  & 1880$\pm$250  270  17 & 1145$\pm$440 220  31 &  500$\pm$168 330  16&  634$\pm$165 425 15& 1100$\pm$170 370 21 \\
168607F   &  B7Ia    &  1.55  & 3100$\pm$290  260  18 &                      &  690$\pm$158 310  14&  970$\pm$205 390 15& 1162$\pm$230 340 26 \\
168625F   &  B6Ia    &  1.47  & 3100$\pm$400  170  16 & 2570$\pm$570 200  35 &  715$\pm$170 290  14& 1037$\pm$259 290 15& 1189$\pm$221 290 24 \\
169454F   &  B4Ia    &  1.00  & 1700$\pm$350  210  14 &                      &  360$\pm$130 340  14&  540$\pm$141 440 15&  550$\pm$168 370 15 \\
179406H   &  B3V     &  0.31  &  420$\pm$90   500  14 &                      &                     &  120$\pm$80  650 12&   70$\pm$43  760 13 \\
183143E   &  B6.5I   &  1.28  & 3100$\pm$390  220  18 & 2530$\pm$230 510  31 &  640$\pm$135 410  15&  760$\pm$180 320 13& 1480$\pm$285 350 22 \\
185859B   &  B0.5Ia  &  0.57  &  710$\pm$130  250  13 &                      &  360$\pm$210 195  14&  495$\pm$232 325 20&  351$\pm$210 270 20 \\
204827B   &  O9.5IV  &  1.07  &  750$\pm$260  175  13 &                      &  350$\pm$150 220  13&  233$\pm$262 230 16&  380$\pm$375 150 22 \\
208501B   &  B8Ib    &  0.63  &  740$\pm$310  160  16 &                      &  305$\pm$145 200  13&                    &                     \\
319703F   &  O6      &  1.50  & 2310$\pm$380  190  17 & 2390$\pm$560 200  38 &  410$\pm$125 300  14& 1128$\pm$293 302 14& 1320$\pm$205 300 26 \\
\enddata
\end{deluxetable}
\end{landscape}
\clearpage
}

\section{Results}

Values of the equivalent widths for DIBs and column densities for K{\sc i} and CH
are presented in Tables 1 and 2.
The equivalent width errors were estimated using the method from
\citet{VE06} in which both spectral noise and uncertainty of the continuum normalization are taken into account.

\begin{figure}
\plotone{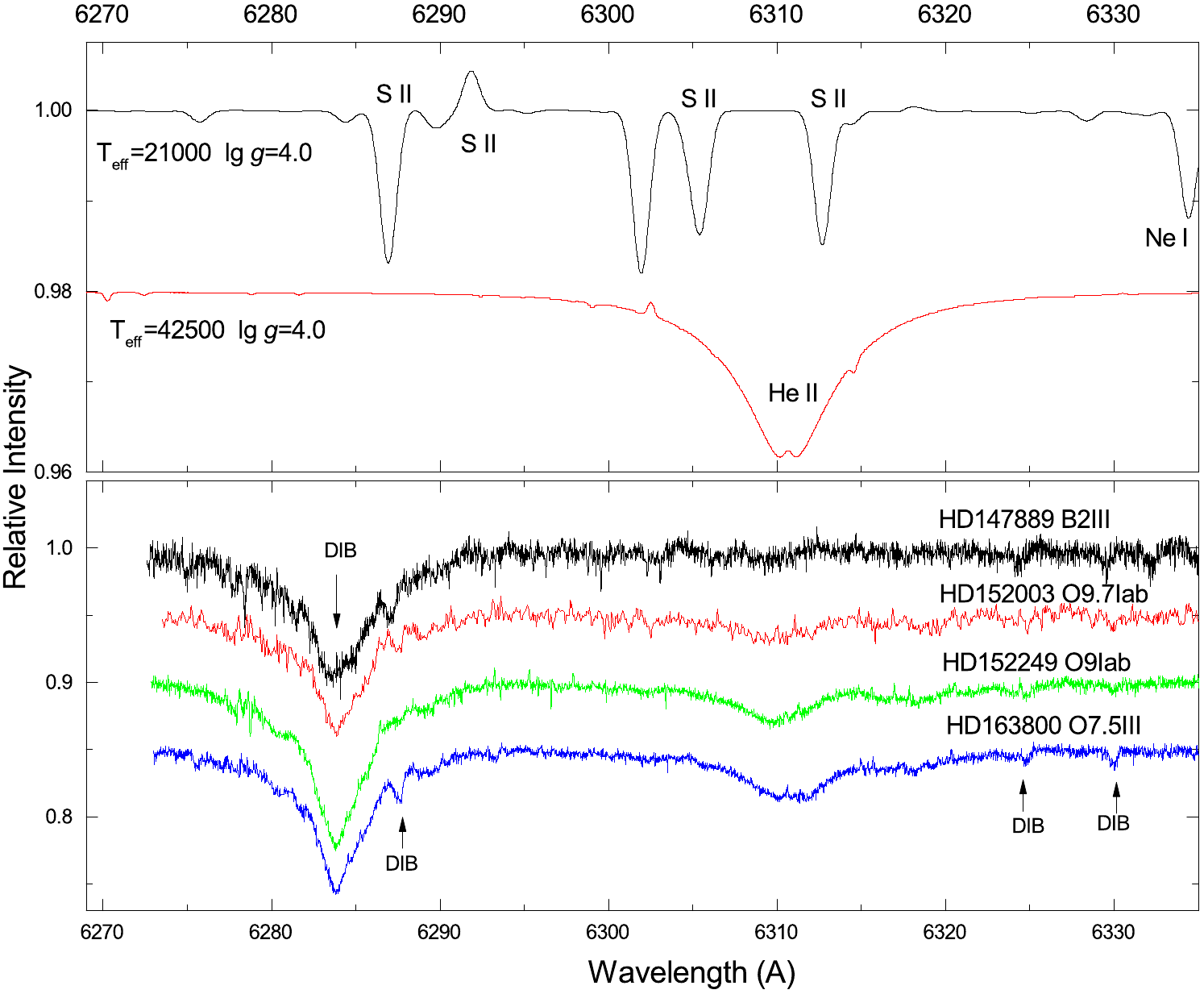}
\caption{The feature mentioned as possible 6311 DIB in work by \citet{SYH18} is, in fact,
a stellar He{\sc ii} line, characteristic for O type stars (see text). As the ion line it is especially strong in supergiant spectra.}
\label{f6311}
\end{figure}

\cite{SYH18} mention many very broad DIBs but only some of them are marked as confirmed.
 Let's consider one example marked by \citep{SYH18} as probable:
the 6311 DIB was originally mentioned by \citep{Her75} as a broad unconfirmed feature seen in HD183143.
However, there are no data for this DIB in the measurements given in table of \citep{Her75}.
Then, in Jeniskens \& Desert (1994) the feature is marked as present, however profiles in Fig. 6 (page 65 in the cited article) hardly prove this statement.
Nevertheless, Tuairisg et al (2000) measured the feature as 6311.53 in HD183143, BD+40\,4220 and BD+63\,1964 though without addressing the stellar and interstellar
contamination effects,  which are severe in this area. Indeed, our spectra of the latter two objects do not
confirm the presence of any broad feature in this area while the available spectra of HD 183143 cannot prove or disprove the presence of 6311.53 DIB
due to the stellar contamination and continuum normalization issues.
In our plot (Fig.~\ref{f6311}) we clearly demonstrate that the suspect DIB is 
a stellar He{\sc ii} line, typically observed in O-type stars. The feature
does not exist in HD147889 as the latter is of B3V Sp/L,  despite a very high reddening. In the upper panel of
Fig.~\ref{f6311} we provide a comparison with the synthetic spectra calculated by \cite{LaHu03} and \cite{LaHu07}
freely available  in the TLUSTY web page\footnote{http://nova.astro.umd.edu/Tlusty2002/tlusty-frames-models.html}.
The shown stellar lines' identification is based on the VALD database \citep{Pis95} line lists compiled with the
effective temperature and logarithm of gravity values indicated for each synthetic spectrum in Figure 2.

The 4882 feature is among the very broad DIBs, mentioned by \citep{Her75}. Its
measurements are naturally difficult because of the close proximity of the
stellar $H_{\beta}$ line. This explains why the number of measurements of
this feature given in Table 1 is substantially reduced.
Worth mentioning is the  rather poor correlation between this diffuse band and the amount of dust particles, which is demonstrated
in Figures 3 and 8.

\begin{figure}
\plotone{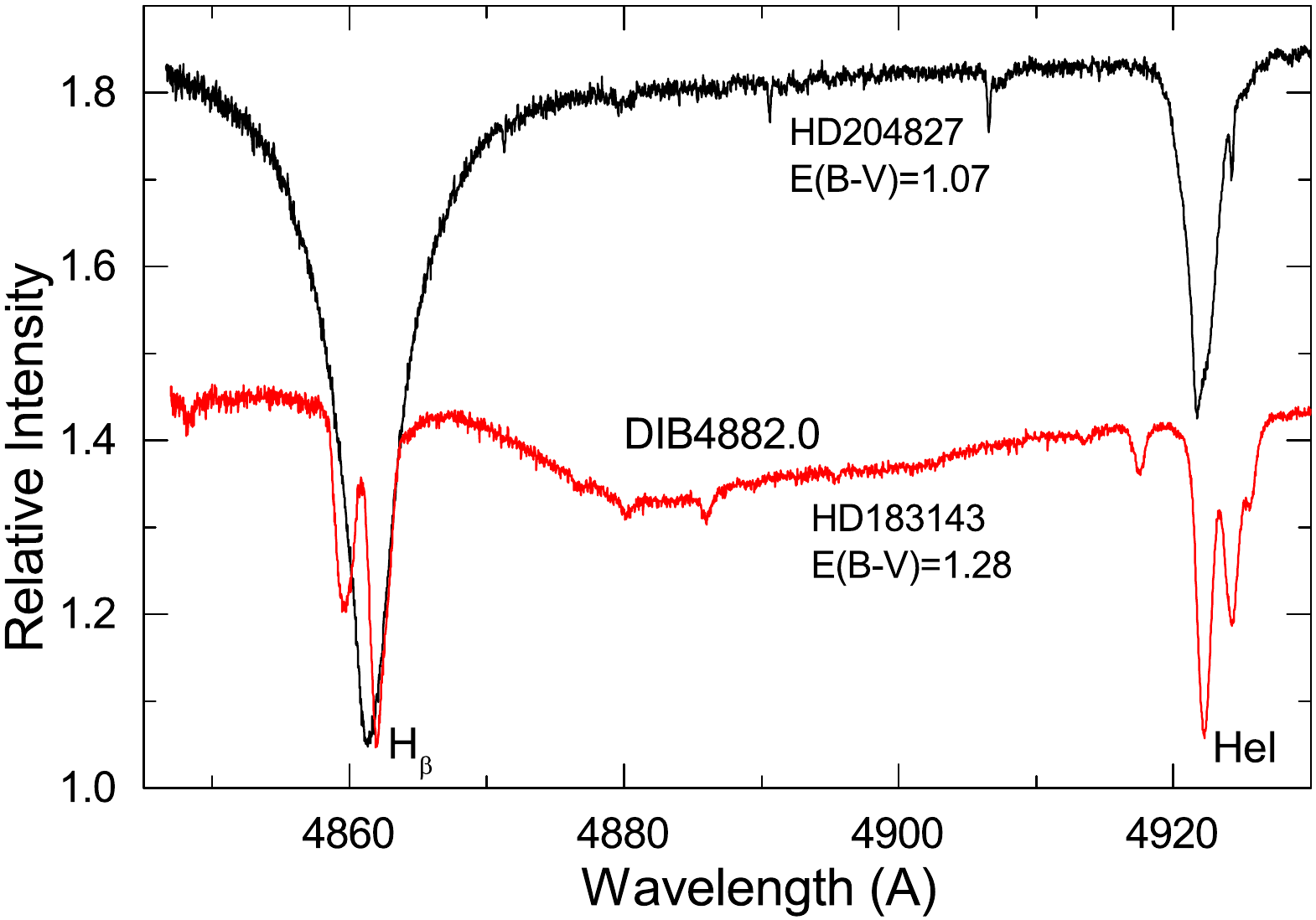}
\caption{Fragments of two spectra of heavily
reddened stars, covering the 4882 very broad DIB. Note the lack of the
simple relation of the DIB to E(B-V).}
\label{d4882}
\end{figure}

\begin{figure}
\plotone{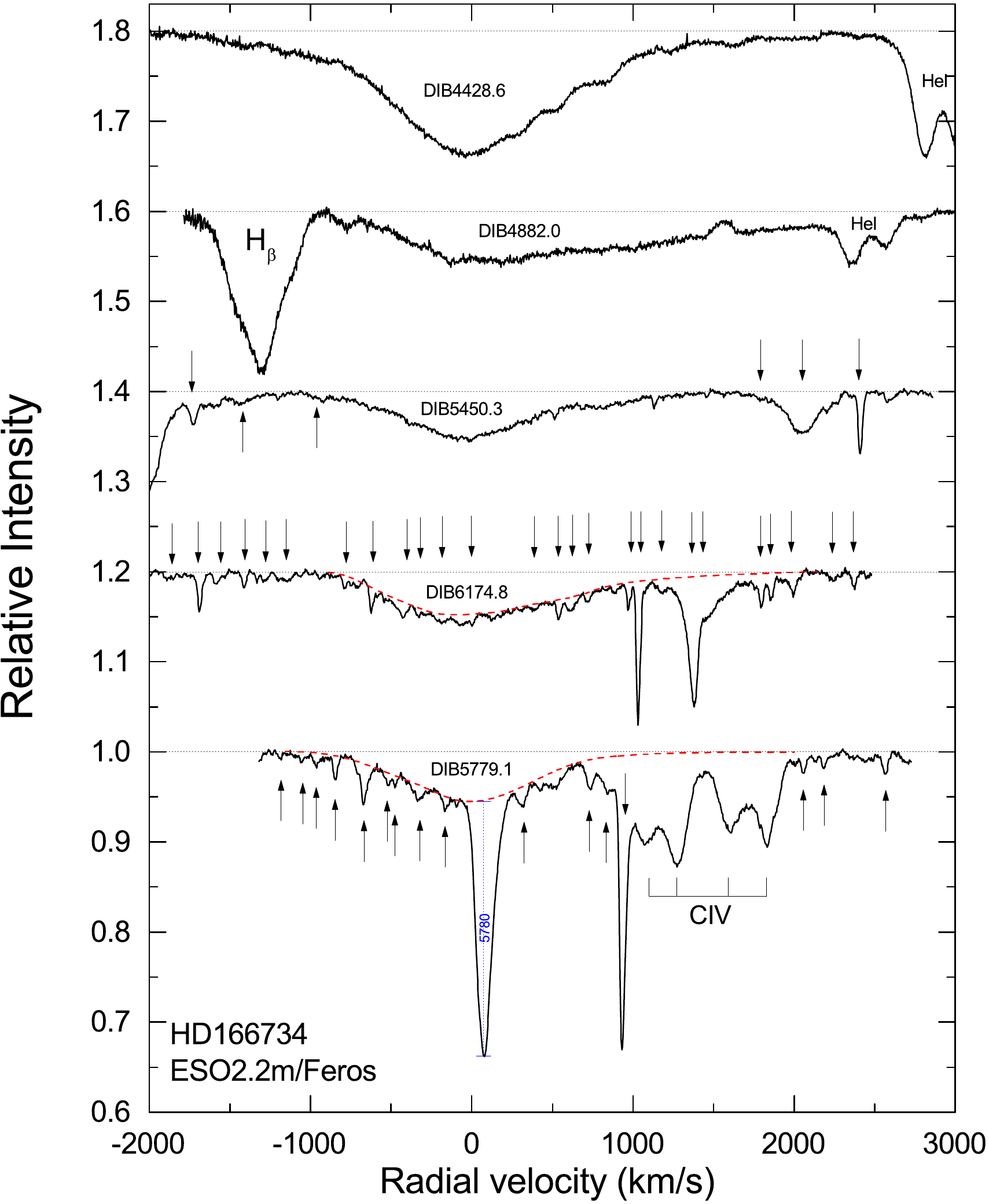}
\caption{Profiles of the confirmed broad DIBs in the spectrum of hot binary O star.
Arrows mark the narrow diffuse bands. Dash lines show profiles of broad DIBs 6175 and 5779.
Note the vertical span of DIB 5780 representing how we measure this feature.
}
\label{hd166734}
\end{figure}

To verify what has been measured in our spectra, we show the
profiles of our confirmed very broad DIBs (Fig. 4). To extract the
profiles we have selected a very heavily reddened O7.5Iab star, HD166734, where
the stellar contaminations in the DIB profiles are the smallest. It is of
importance to mention that, e.g. the profile of the 6175 DIB, presented by
\cite{SYH18} in the spectrum of B7 supergiant HD183143, is likely shown as too
broad, perhaps because of stellar contaminations in its blue side. In all our
spectra of O-type stars the profile begins around 6155~\AA, not around
6130~\AA. This is the possible source of errors: the borders of very broad
features are ill-defined. It is thus very important to compare the spectra of
different spectral types taken using different instruments.

Our measurements of the sample of 43 high-resolution, high-S/N ratio
spectra do confirm the already reported (see the Introduction) correlations
of the very broad DIBs with a color excess;  Figs. 7, 8
present the least-squares linear fits and the correlation coefficients for
broad DIBs. All calculations were performed with the Y-errors taken into account.

We have measured the EWs of the broad DIBs (see
Table 1), fitting the profile manually to separate the
stellar contaminations (\cite{Her66}; \cite{KWGH87}), especially
strong in B type stars. Such a procedure leaves some arbitrariness but
a more precise method hardly exists as the stellar spectra are a bit
unpredictable (see an example of measurements in Fig. 5). This may lead to a correlation coefficient a bit
smaller than the true one. In any case the broad features 4430, 6175 and 5779 do
correlate quite tightly with E(B$-$V) while 4882 and 5450 demonstrate a lower
magnitude of correlation with the amount of interstellar dust.

\begin{figure}
\plotone{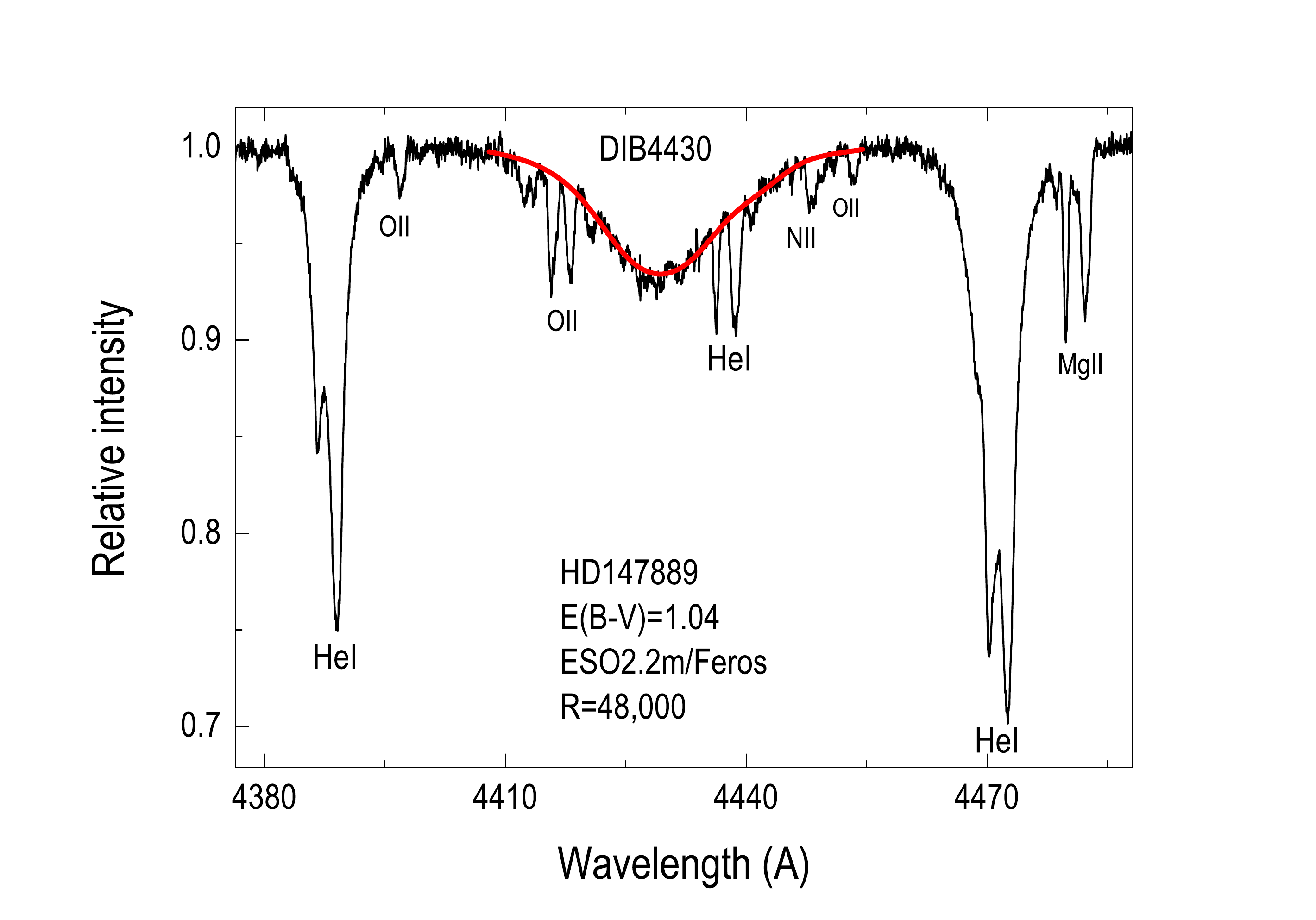}
\caption{Example of the measurement of the broad (FWHM 18 \AA) DIB4430 with the manually fitted profile overplotted on the original spectrum.}
\label{d4430-147889}
\end{figure}

Some diffuse bands are severely contaminated  by telluric lines, e.g. reasonably broad DIB6284.
Fortunately, spectra from fiber-fed spectrographs provide almost excellent removal of telluric lines
by means of telluric standard stars. A good example is given in Fig. 6.

\begin{figure}
\plotone{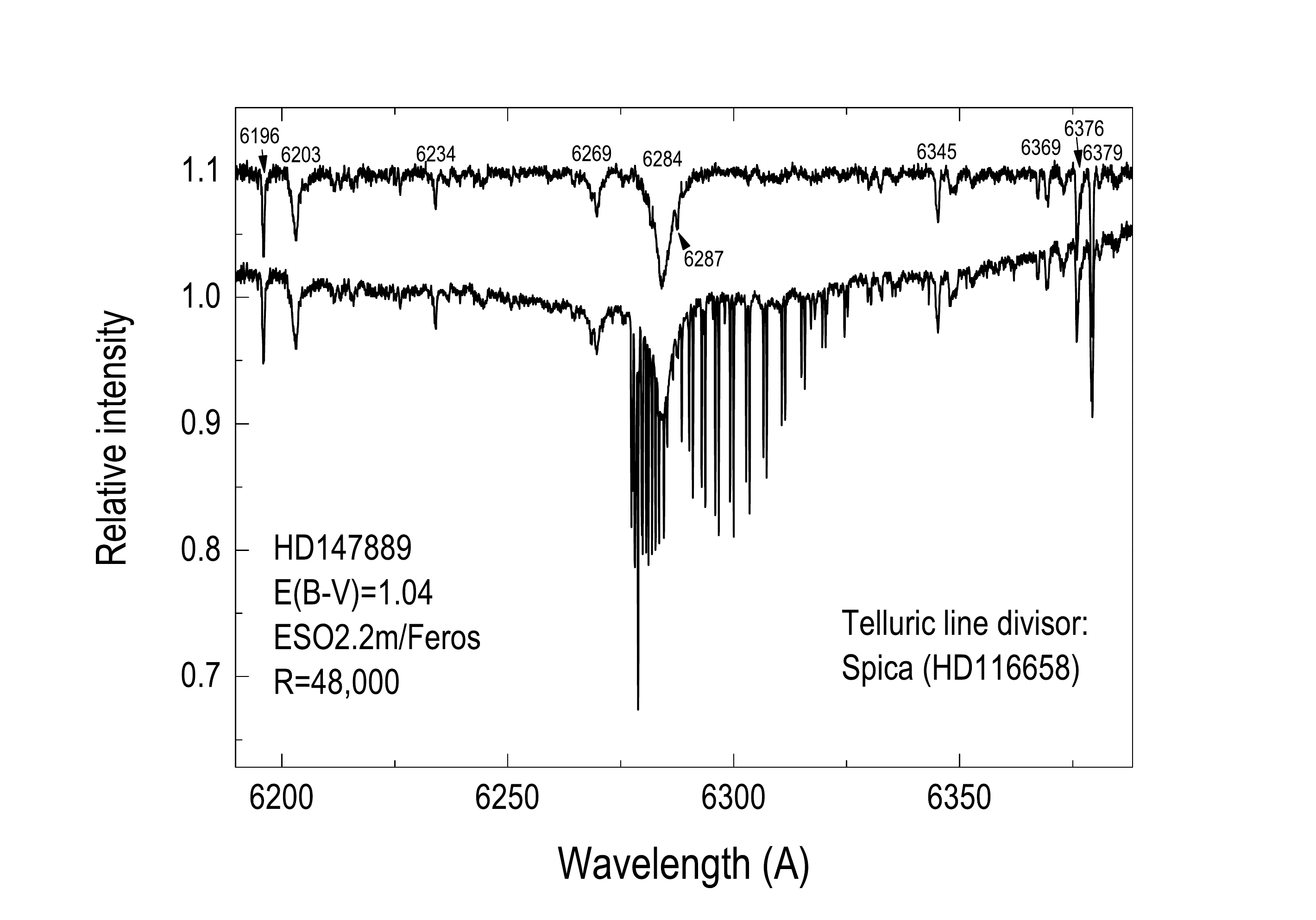}
\caption{Example of telluric lines removal
from the profile of broad DIB 6284. Narrow DIBs are also marked. The lower spectrum is a raw data just normalized to just one point for clarity.}
\label{tellur}
\end{figure}

One of the very broad features, depicted at Fig. 4, is the
5450 DIB. It is reasonably weak and its profile is contaminated with
stellar lines in B-type stars. The above makes the measurements of
EWs uncertain. This is why a correlation between this DIB and
\ce\ is lower than that of DIBs 4430, 6175, and 5779 (see Fig. 7).


\begin{figure}
\centering
\includegraphics[width=10 cm, angle=270, clip=]{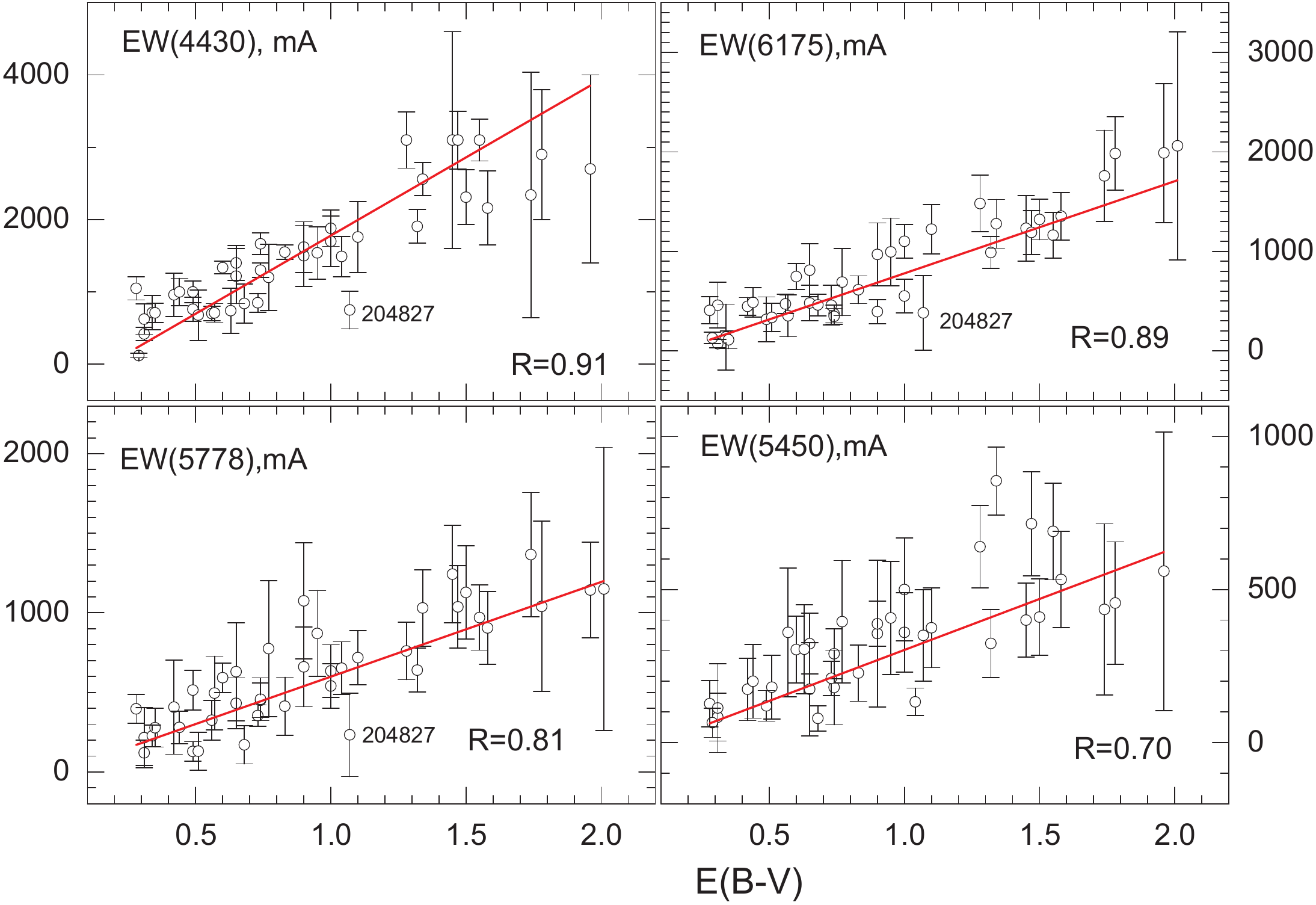}
\figcaption{Weighted linear fits with reasonably tight correlations between
EWs of four very broad DIBs and \ce.}
\label{DIBvsCE}
\end{figure}

\begin{figure}
\centering
\includegraphics[width=10 cm, angle=270, clip=]{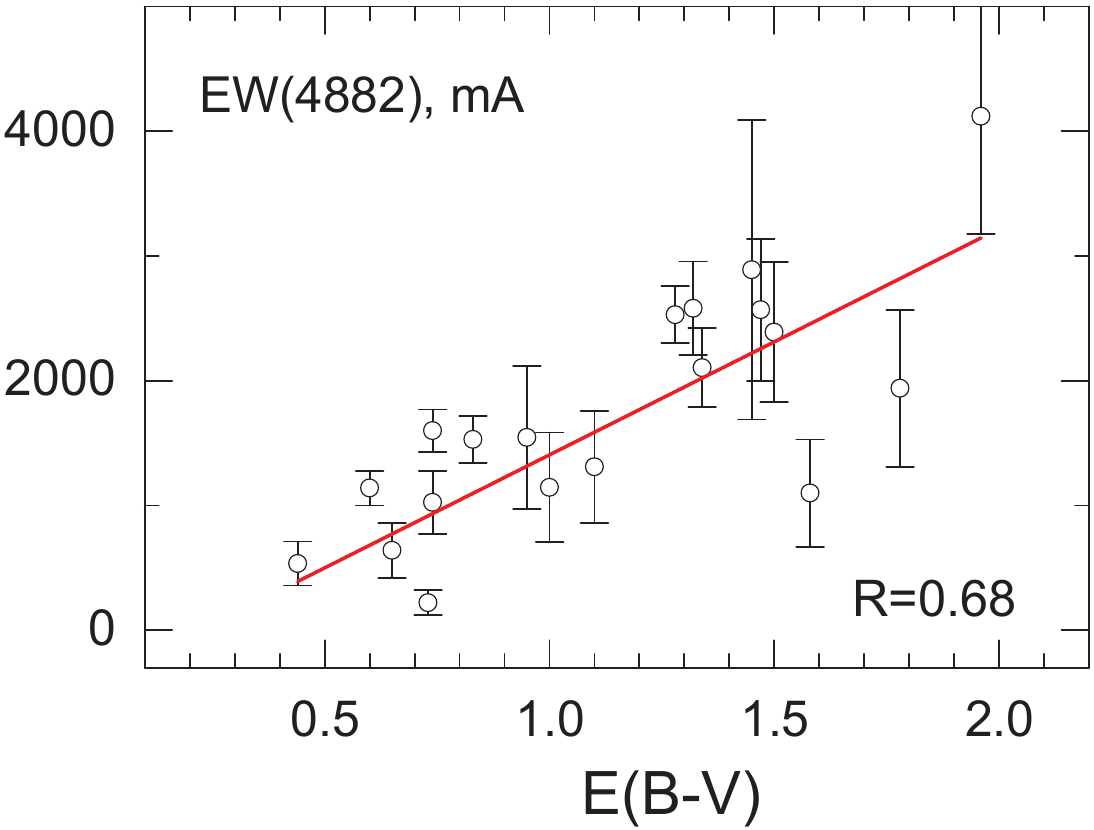}
\figcaption{Weighted linear fits between EWs of 4882 DIB and \ce.}
\label{DIB4882vsCE}
\end{figure}

A closer look into the area around 6175 reveals that the
observed broad feature is a blend of at least two also broad bands centered approximately at 6170 and 6177~\AA.
Sometimes this separation is quite evident due to deep 6170 DIB: Cyg OB2 objects, HD 152233, HD 154368, HD 163800, etc.
However, e.g.  BD-13 4923, BD-13 4928, BD-13 4929, HD73882, HD319703, etc.  exhibit a rather smooth
profile without any indications of the presence of the blend.

It is particularly surprising that very broad 6175 DIB is pretty tightly correlated with E(B$-$V)
(Fig. 7). Despite a possible source of uncertainty, due to
the manual fit and the abovementioned issues, the correlation coefficient is as high as  0.89. Thus
the carriers of the very broad DIBs seem to be well mixed with other
interstellar species, including dust grains. The correlation, based on
our sample of 40 high-resolution, high-S/N ratio spectra, looks
tighter than those presented by \cite{SYH18}. The correlation would be even tighter
but HD204827 departs down from the average relation and this effect is certainly not due to a measurement error.

\cite{SYH18} tried to relate the very broad diffuse band 6177 (we mark it as 6175) to other, narrower
ones, in particular DIB 5780. The latter is the
major one, discovered by \cite{Heg22}. Both features in the Figure 10 of \cite{SYH18} seem to
be closely related but the observed scatter seems to exceed the
measurements' errors.

\begin{figure}
\centering
\includegraphics[width=20 cm]{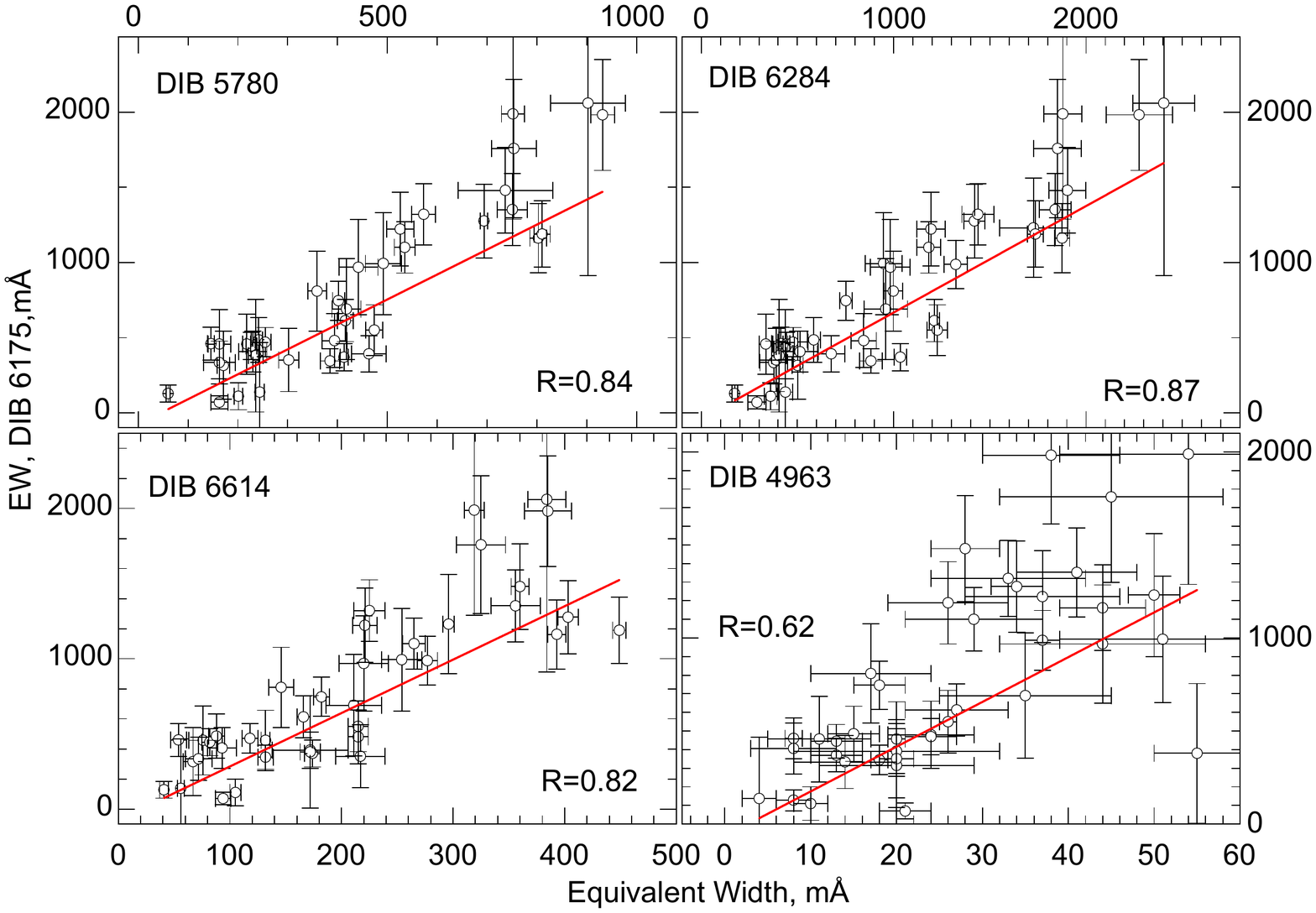}
\figcaption{Correlation of broad 6175 DIB with other, narrower diffuse bands with tight
correlation between the very broad 6175 DIB and the much narrower 5780 and 6284
ones. Apparently their carriers are to be found in the same
environments.}
\label{DIBvsDIB}
\end{figure}

We have also checked the interrelations between very broad and
narrower DIBs. The weighted linear fits with Pearson's correlation
coefficients are depicted on Fig. 9.  The correlations'
magnitudes were calculated with measurement errors taken into account as the weights. All the fits
presented on the Fig. 9 have the slope's standard error
between 0.07 and 5.05: with their best value (0.07) for the relation 6175 vs 6284 DIB and the
worst one (5.05) for the relation 6175 versus 4963 DIB.

It is evident that the blend of broad DIBs at 6175 correlates very well with DIBs 5780,
6614 and, especially with 6284, which supposes a similar physical
environment for their carriers. Its correlation with the narrow 4963
DIB is not so tight. It is, however, much more evident than that
presented by \cite{SYH18}. Let's emphasize that our sample is twice
as big as that in the latter paper. Anyway, the mutual correlations between
the abovementioned DIBs are never as tight as that between 6614 and 6196
DIBs \citep{KGBB16}.

\citet{SYH18} divided the observed DIBs into the sets related to
either atomic or molecular gas. The last one was considered as the environment
for the 4963 DIB carrier. To verify this we related the column
densities of the K{\sc i} 7699~\AA\ line and the CH 4300.3~\AA\ molecular
feature to this band strength. Weighted fits, shown in
Fig. 10, have the slope's standard error 0.06 for K column
densities versus 4963 EWs and 0.19 for CH column densities versus 4963 EWs. As
seen in Fig. 10 the 4963 DIB correlates better with
the atomic feature than with the molecular one. The former would be
much better if not the specific object --- HD147889. It is hard to say why
the K{\sc i} line is so weak in the spectrum of this star. It was, however, verified
in a few spectra from different instruments and thus is real.  Anyway
Fig. 10 creates some doubts on whether the 4963 carrier is
really situated in molecular gas rather than in atomic gas.
On the other hand, Thorburn et al. (2003) marked DIB 4963 as so-called C$_2$-DIB correlating
with the abundance of the interstellar C$_2$ molecule. However, this conclusion was questioned by Galazutdinov et al.(2006)
where the authors estimated the correlation coefficient as low as 0.56 for 20 measurements. Recently,
Elyajouri et al. (2018) estimated the correlation between intensities of 18 C$_2$-DIB including DIB 4963
and reported the correlation level for it as high as 0.95. The estimation is based on a rather small number
of measurements thus the significance of the attribution of DIB 4963 as a C$_2$-DIB remains low.

In Fig. 11 we provide the general average of all measured profiles of broad diffuse bands with estimated
FWHM and the rest interstellar wavelength position. A good question is: where is the center of very asymmetric DIB4882?
As it is shown in Fig. 11, we attributed the rest wavelength position to the deepest part of the feature's profile, namely at
4882.0 \AA. Thus, this diffuse band is a holder of a very extended right (red) wing with a rather sharp cut on the left (blue) wing that mimics the behavior of well-known
bands of simple molecules. The average FWHM of this band, 32 \AA\, makes it the broadest feature of the sample.
Other bands' central wavelengths and FWHM (given in the parentheses) are 4428.6 (17 \AA), 5450.3 (14 \AA), 5779.1 (16 \AA) and 6174.8 (25 \AA).
The variability of the FWHM from object to object (Table 1) can be explained by the following reasons:
(i) due to the shallowness of the broad features, the FWHM is very sensitive to the continuum normalization uncertainties and the position of the deepest point of the profile.
The situation is more complex if the S/N ratio is low and in cases of severe stellar contamination. (ii) The FWHM may grow with the increasing the number of
populated transitions of carrier molecule, i.e. the variability of the FWHM may be of physical origin.

\begin{figure*}
\centering
\includegraphics[width=6 cm, angle=270]{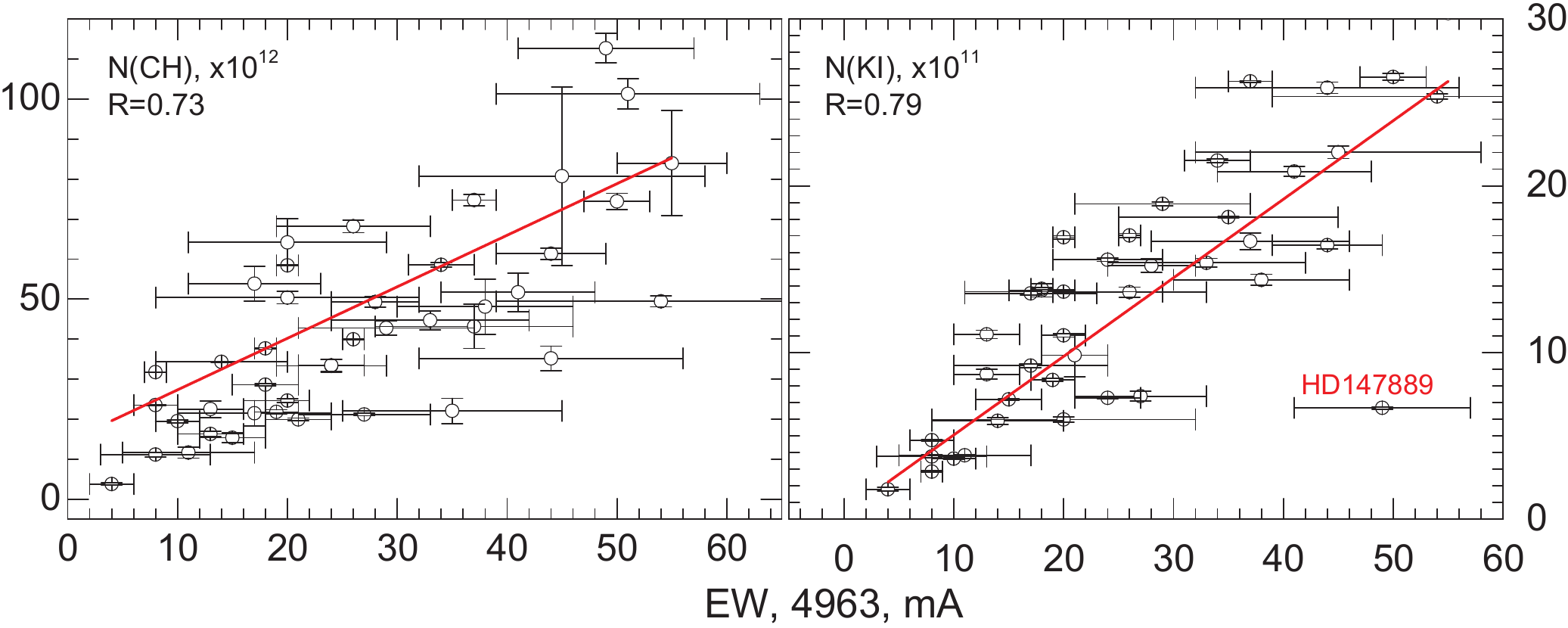}
\figcaption{Correlation of DIB 4963 with CH and K. There is no evidence that the DIB
originates in molecular clouds.}
\label{DIB4963}
\end{figure*}

\begin{figure*}
\plotone{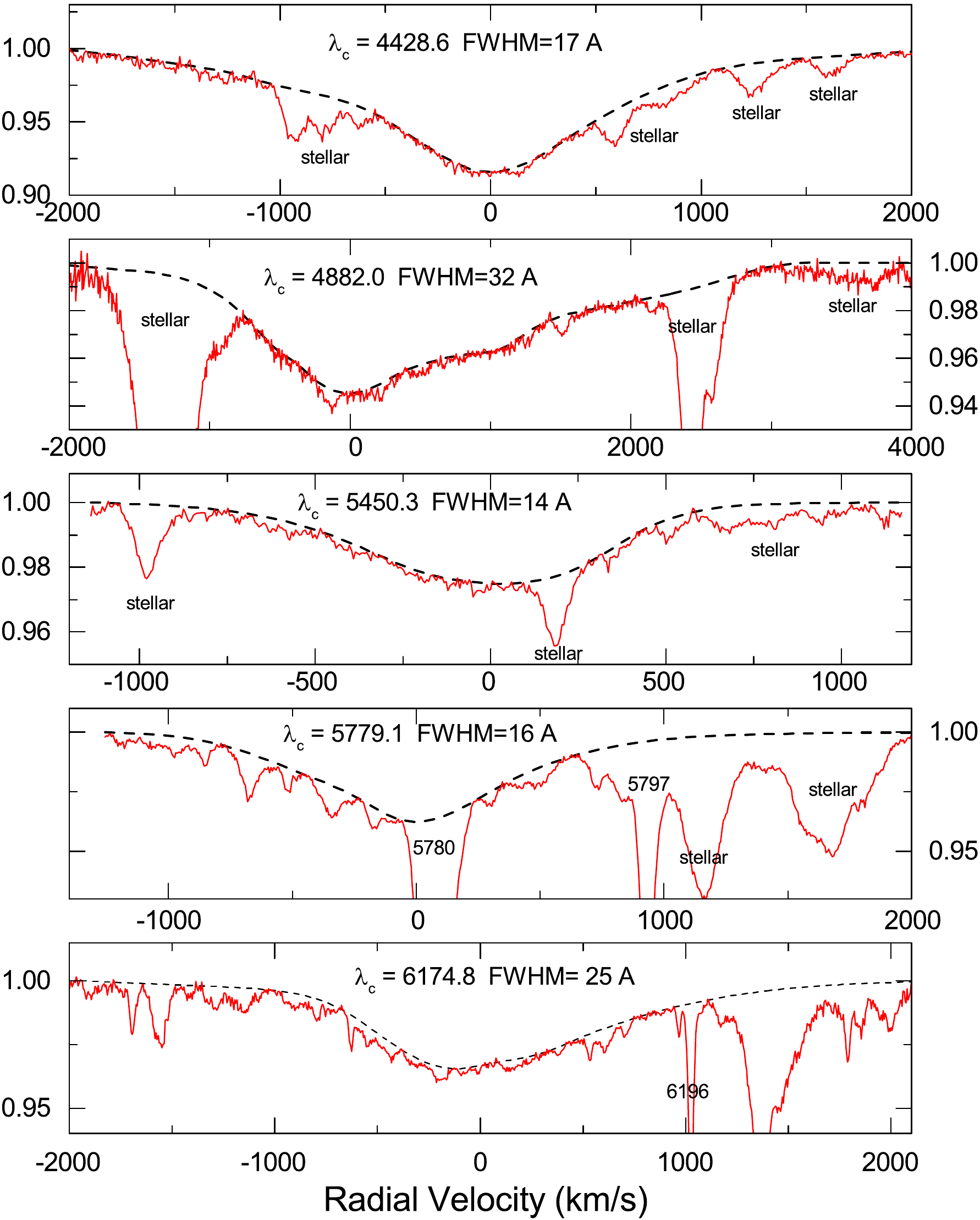}
\caption{General average of the profiles of studied DIBs in the interstellar radial velocity scale with estimated rest wavelength
and FWHM. All unmarked absorptions are diffuse bands; see Fig. 4.
}
\label{generalprof}
\end{figure*}

\clearpage
\begin{deluxetable}{lrrrrrr}
\tabletypesize{\footnotesize}
\tablewidth{0pt}
\tablecaption{Equivalent widths (m\AA) of some relatively narrow DIBs,
including weak band at $\lambda$ 4963 \AA\ one of so-called C$_2$ DIBs \citep{Th03},
 and column densities (cm$^{-2}$) of CH and K{\sc i}.
\label{ndibs}}
\tablehead{
	\colhead{Star}     &
          \colhead{EW(4963)} &
	                  \colhead{EW(5780)} & \colhead{EW(6284)} &
	                                              \colhead{EW(6614)} & \colhead{N(CH)$\times$E12} &
	                                                                                          \colhead{N(K{\sc i})$\times$E11}
}
\startdata
BD+404220 &  54$\pm$15   &  752$\pm$23   &  1879$\pm$99    & 319$\pm$9    & 49.44$\pm$1.34   &  25.36$\pm$0.15   \\
BD-134923 &  37$\pm$9    &  526$\pm$27   &  1194$\pm$74    & 221$\pm$11   & 43.19$\pm$5.49   &  16.68$\pm$0.50   \\
BD-134927 &  51$\pm$12   &  492$\pm$36   &   948$\pm$95    & 254$\pm$18   & 101.3$\pm$3.80   &  30.99$\pm$0.35   \\
BD-134928 &  35$\pm$10   &  418$\pm$29   &   959$\pm$113   & 211$\pm$25   & 22.08$\pm$3.16   &  18.12$\pm$0.05   \\
BD-134929 &  44$\pm$12   &  442$\pm$39   &   983$\pm$103   & 220$\pm$22   & 35.21$\pm$3.06   &  25.88$\pm$0.34   \\
CygOB\_7  &  45$\pm$13   &  754$\pm$45   &  1852$\pm$124   & 325$\pm$22   & 80.74$\pm$22.29 &   22.01$\pm$0.37   \\
CygOB\_8  &  41$\pm$7    &  751$\pm$30   &  1839$\pm$82    & 356$\pm$22   & 51.70$\pm$4.80  &   20.86$\pm$0.30   \\
CygOB11   &  38$\pm$8    &  932$\pm$24   &  2276$\pm$172   & 385$\pm$21   & 48.11$\pm$6.90   &  14.36$\pm$0.33   \\
CygOB12   & \nodata      &  903$\pm$75   &  2404$\pm$160   & 384$\pm$17   &  \nodata         &  46.47$\pm$0.34   \\
Hersch36  &  20$\pm$12   &  463$\pm$35   &   676$\pm$71    & 172$\pm$34   & 50.40$\pm$1.56   &   5.99$\pm$0.17   \\
 15785    &  24$\pm$5    &  394$\pm$25   &   845$\pm$65    & 215$\pm$9    & 33.35$\pm$1.58   &  15.56$\pm$0.10   \\
 24912    &  11$\pm$6    &  163$\pm$23   &   438$\pm$12    &  76$\pm$13   & 11.67$\pm$1.39   &   3.84$\pm$0.03   \\
 34078    &  20$\pm$9    &  171$\pm$12   &   501$\pm$29    &  67$\pm$9    & 64.20$\pm$5.88   &  13.65$\pm$0.10   \\
 73882    &   8$\pm$1    &  146$\pm$7    &   414$\pm$48    &  54$\pm$7    & 31.75$\pm$0.11   &   2.86$\pm$0.04   \\
 76341    &  14$\pm$6    &  163$\pm$32   &   377$\pm$71    &  72$\pm$12   & 34.33$\pm$0.15   &   5.90$\pm$0.21   \\
 78344    &  37$\pm$2    &   \nodata     &  1323$\pm$61    & 277$\pm$9    & 74.78$\pm$1.43   &  26.26$\pm$0.05   \\
 80077    &  50$\pm$3    &   \nodata     &  1727$\pm$177   & 296$\pm$5    & 74.44$\pm$2.04   &  26.53$\pm$0.21   \\
147165    &   4$\pm$2    &  244$\pm$9    &   438$\pm$38    &  56$\pm$3    &  3.81$\pm$0.27   &   1.80$\pm$0.10   \\
147888    &  19$\pm$2    &  248$\pm$22   &   267$\pm$87    &  83$\pm$13   & 21.84$\pm$0.57   &   8.35$\pm$0.09   \\
147889    &  49$\pm$8    &  371$\pm$13   &   435$\pm$22    & 179$\pm$9    &112.70$\pm$3.71   &   6.68$\pm$0.07   \\
148379    &  13$\pm$3    &  413$\pm$8    &  1035$\pm$32    & 174$\pm$7    & 22.53$\pm$2.06   &  11.09$\pm$0.25   \\
149038    &   8$\pm$5    &  230$\pm$28   &   513$\pm$96    &  93$\pm$13   & 11.16$\pm$0.60   &   3.76$\pm$0.09   \\
149404    &  18$\pm$3    &  402$\pm$12   &   752$\pm$33    & 182$\pm$7    & 28.63$\pm$0.32   &  13.73$\pm$0.40   \\
148937    &  17$\pm$7    &  359$\pm$19   &   998$\pm$48    & 146$\pm$11   & 21.47$\pm$3.18   &   9.21$\pm$0.11   \\
149757    &   8$\pm$2    &   60$\pm$4    &   174$\pm$15    &  41$\pm$4    & 23.52$\pm$0.04   &   4.74$\pm$0.04   \\
152233    &  13$\pm$3    &  227$\pm$11   &   444$\pm$40    &  82$\pm$7    & 16.30$\pm$0.65   &   8.70$\pm$0.30   \\
152235    &  18$\pm$1    &  385$\pm$19   &   882$\pm$59    & 132$\pm$7    & 37.70$\pm$0.20   &  13.82$\pm$0.07   \\
152249    &  15$\pm$3    &  242$\pm$8    &   585$\pm$34    &  88$\pm$7    & 15.36$\pm$1.22   &   7.19$\pm$0.05   \\
154368    &  20$\pm$1    &  218$\pm$8    &   336$\pm$36    & 132$\pm$4    & 58.45$\pm$0.10   &  16.91$\pm$0.10   \\
154445    &  10$\pm$2    &  201$\pm$9    &   361$\pm$33    & 105$\pm$5    & 19.53$\pm$0.39   &   3.64$\pm$0.06   \\
157038    &  27$\pm$6    &  416$\pm$16   &  1212$\pm$27    & 166$\pm$6    & 21.20$\pm$0.31   &   7.38$\pm$0.30   \\
163800    &  24$\pm$3    &  255$\pm$12   &   477$\pm$78    & 118$\pm$7    & 33.46$\pm$1.48   &   7.28$\pm$0.05   \\
166734    &  34$\pm$3    &  694$\pm$7    &  1420$\pm$54    & 403$\pm$9    & 58.60$\pm$0.52   &  21.51$\pm$0.11   \\
168112    &  29$\pm$8    &  535$\pm$21   &  1183$\pm$62    & 265$\pm$10   & 42.80$\pm$1.77   &  18.92$\pm$0.11   \\
168607    &  44$\pm$5    &  803$\pm$17   &  1876$\pm$39    & 393$\pm$8    & 61.36$\pm$1.41   &  16.44$\pm$0.24   \\
168625    &  26$\pm$7    &  811$\pm$16   &  1737$\pm$41    & 449$\pm$6    & 68.24$\pm$1.58   &  13.62$\pm$0.30   \\
169454    &  26$\pm$1    &  474$\pm$17   &  1227$\pm$53    & 215$\pm$9    & 39.96$\pm$0.11   &  17.02$\pm$0.13   \\
179406    &  21$\pm$3    &  163$\pm$17   &   290$\pm$48    &  94$\pm$7    & 19.95$\pm$0.36   &   9.84$\pm$1.30   \\
183143    &  28$\pm$4    &  737$\pm$95   &  1903$\pm$95    & 360$\pm$8    & 49.30$\pm$1.29   &  15.21$\pm$0.40   \\
185859    &  20$\pm$2    &  302$\pm$20   &   390$\pm$89    & 217$\pm$22   & 24.67$\pm$0.52   &  11.03$\pm$0.10   \\
204827    &  55$\pm$5    &  236$\pm$15   &   405$\pm$39    & 172$\pm$9    & 84.00$\pm$13.10  &  30.35$\pm$0.20   \\
208501    &  17$\pm$6    &  248$\pm$38   &   524$\pm$95    & 129$\pm$12   & 53.86$\pm$4.36   &  13.55$\pm$0.12   \\
319703    &  33$\pm$9    &  573$\pm$24   &  1438$\pm$83    & 225$\pm$14   & 44.74$\pm$2.34   &  15.38$\pm$0.26   \\
\enddata
\end{deluxetable}
\clearpage

\section{Discussion}

Our conclusions do not always agree with those of \cite{SYH18}. In many
cases we can confirm the latter:  most of the mutual correlations
between different DIBs look very similar. Thus we confirm the existence
of strong correlation between a broad 6175 DIB and two medium 5780
and 6284 ones where the Pearson correlation coefficients are 0.84 and
0.87, respectively. This may confirm that they have
the same molecular carrier \citep{MC10}.  It is worth mentioning  that
our paper \citep{KGBB16} demonstrates examples of evidently different
6196 vs. 6614 strength ratios and thus it is difficult to
state that any two DIBs are of common origin.

However, we do not confirm a poor correlation (between the 6175 DIB and
\ce (r=0.57) of \cite{SYH18} versus r=0.89 in the present paper, as well as too low of a degree
of correlation (r = 0.01 in \cite{SYH18}) between 6175 and 4963 DIBs,
it is 0.62 in our case.

It is also rather risky to divide the observed DIBs in between of
atomic and molecular gas clouds. It is clear from Fig. 10
for the case of the relatively weak and narrow diffuse band at
$\lambda$~4963~\AA, so-called C$_2$ DIB. The figure demonstrates
a better relation between this DIB and the atomic K{\sc i}, not the CH molecule, as
it could be inferred, e.g. from the \cite{SYH18}, where, on the basis
of poor correlation between the 6175 and 4963 diffuse bands, the authors
suggest an atomic gas environment for DIB 6175.

The current list of the observed molecules in ISM is available at
\url{https://www.cv.nrao.\\ edu/\~{}awootten/allmols.html}. The most numerous
group among them are two- and three-atomic molecules. An attempt to fit
vibrational contours of the molecules to the very broad DIBs results in
the following assumptions:

\begin{enumerate}

\item 4430 $\rightarrow$ CuCl D$^1\Pi$--X$^1\Sigma$ (0, 1) medium-strong band with a red-degraded double head at 4433.8; 4881 $\rightarrow$ CuCl
B$^1\Pi$--X$^1\Sigma$ (0, 0) strong band with a red degraded head at 4881.5. It is
very interesting that the CuCl bands occur rather frequently as
impurities in other spectra, especially in flames, fluorescence,
absorption, discharge tubes, and also in arc. They also appear when
CuCl is introduced into active nitrogen \citep{RAB62}; however, the
very low abundances of both elements make this hypothesis uncertain.

\item 5450 $\rightarrow$ BO$_2$, headless and narrow band (maximum
intensity at 5450). The waves of the bands are observed when boric acid is
introduced into an arc or flame, or when finally divided boron is
burnt \citep{KM61}; once again the abundance of boron makes the
hypothesis unlikely.

\item 5779 $\rightarrow$ SrF, yellow system
(B$^2\Sigma^+$--X$^2\Sigma^+$) (0, 0) band with a close double heads
at 5779.5 in $Q$ branch and 5772.0 in $R$ branch \citep{NG67}; it also
looks like a chance coincidence of spectral features since the
abundances of strontium and fluor are very rare in the space;

\item 6175 $\rightarrow$ FeO, orange (A0, 2ii) strong band of A system
(red degraded head at 6180.5) \citep{PG84}. In this case both
elements are quite abundant which makes this hypothesis much more
likely than the former ones.

\end{enumerate}

Among them, only the existence of the FeO molecule has been confirmed
in the ISM (\url{https://cdms.astro.uni-koeln.de/cdms/portal/}). Also, the abundance of oxygen and iron is quite high
in comparison to rather exotic other elements listed in the first three cases. Thus only the last item
may have some real meaning. Thus the molecular spectroscopy  in its present state gives us
no suggestion as to what the carriers of broad DIBs may be.

\acknowledgments
J.K., R.H. and W.S. acknowledge the financial support of the Polish National Science
Centre under the grant 2017/25/B/ST9/01524 for the period 2018
\textendash\ 2021. G.G. and J.K. acknowledge the financial support of the
Chilean fund CONICYT grant REDES 180136.

\clearpage

\appendix

\section{Appendix}

\subsection{Profiles of 4430 DIB}

\includepdf[pages={1-6}]{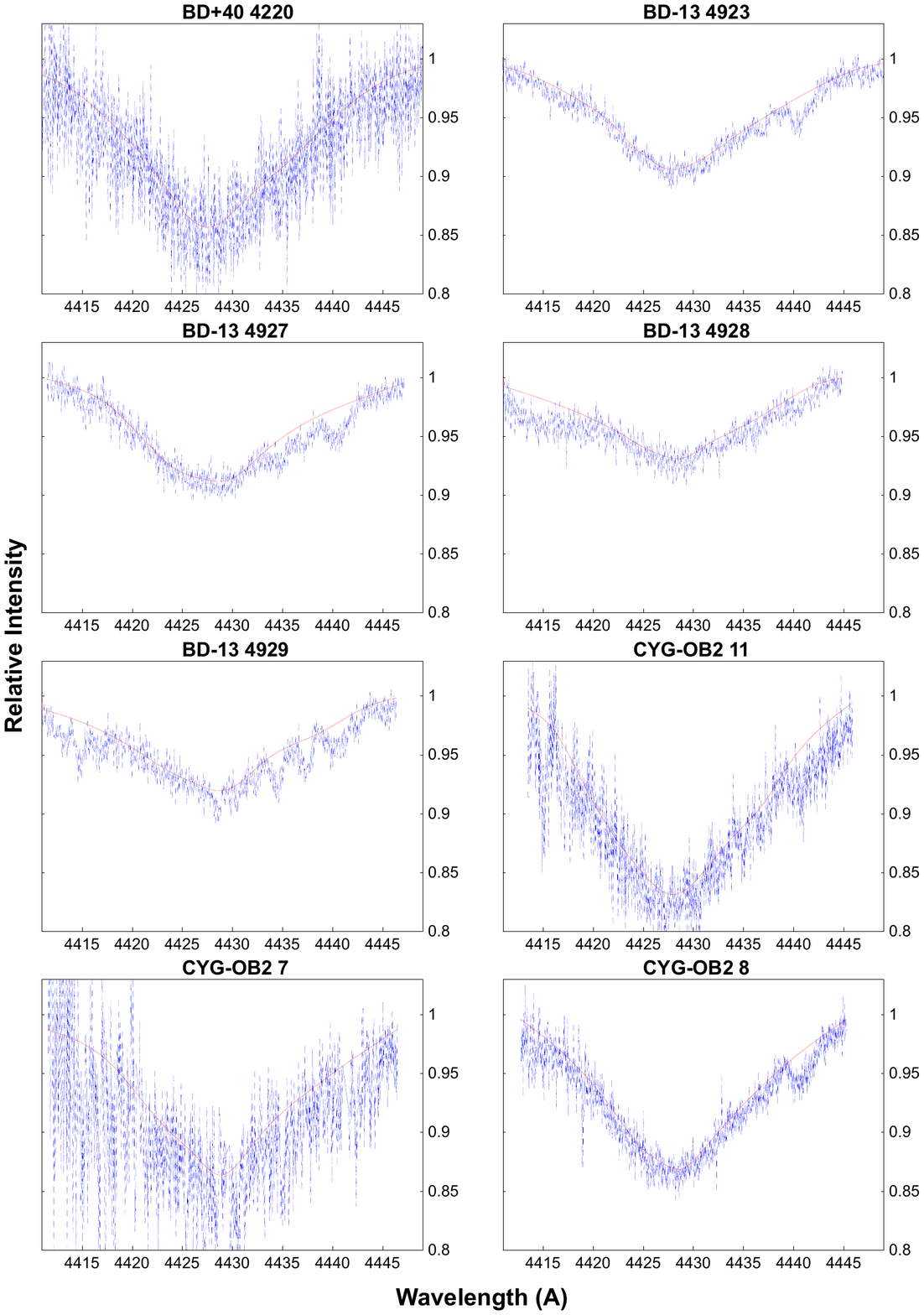}

\subsection{Profiles of 4882 DIB}

\includepdf[pages={1-3}]{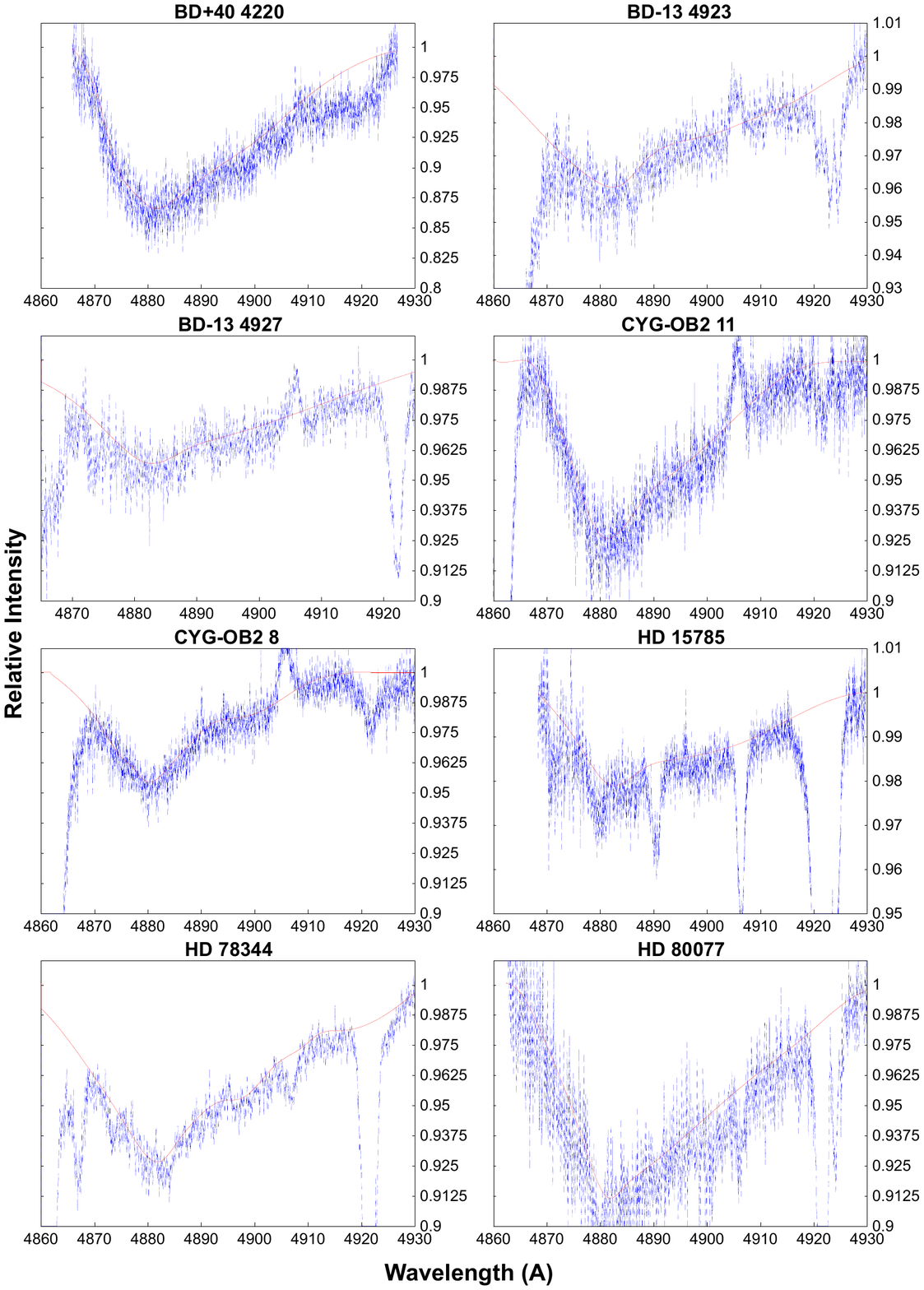}

\subsection{Profiles of 5450 DIB}

\includepdf[pages={1-5}]{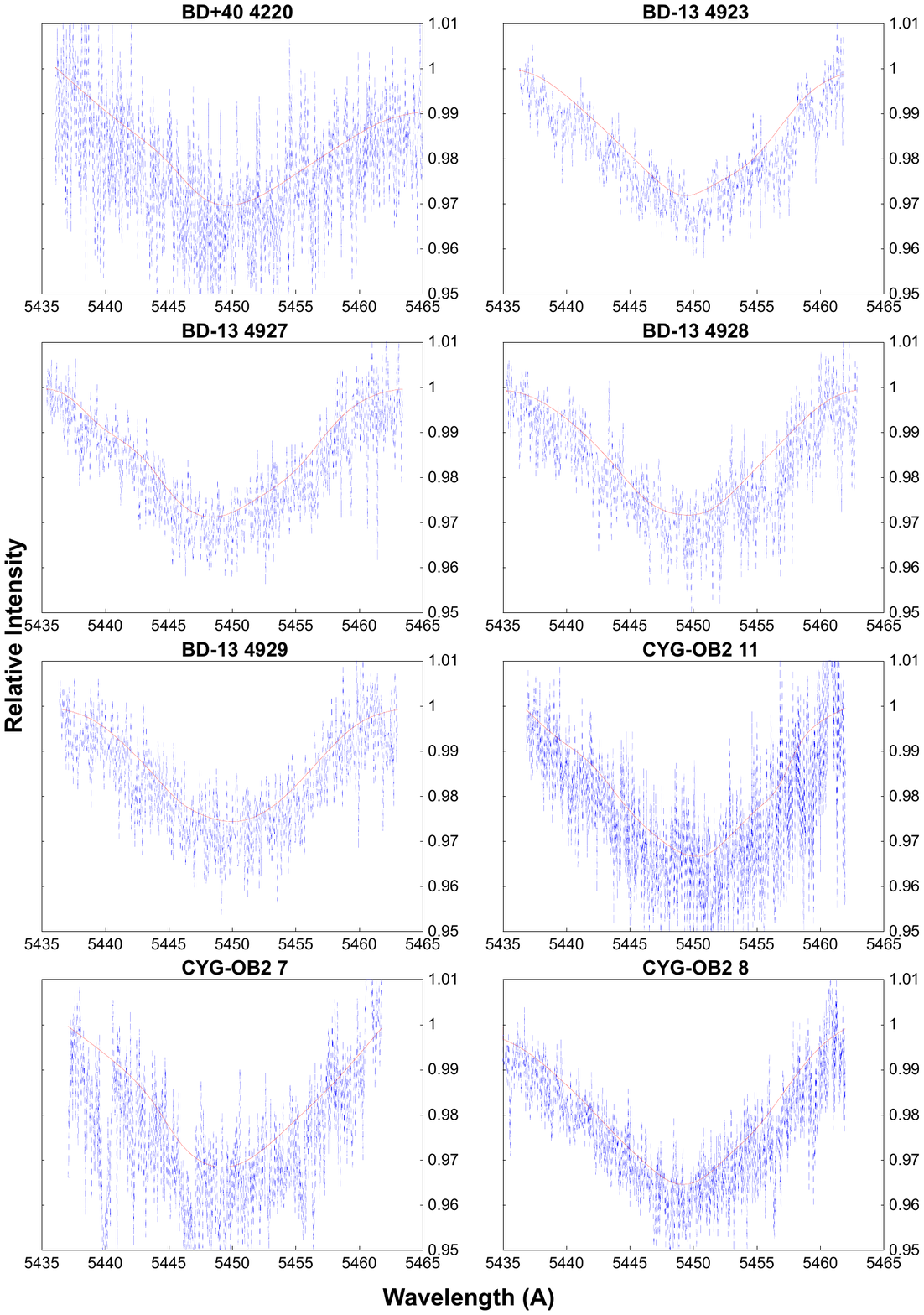}

\subsection{Profiles of 5779 DIB}

\includepdf[pages={1-6}]{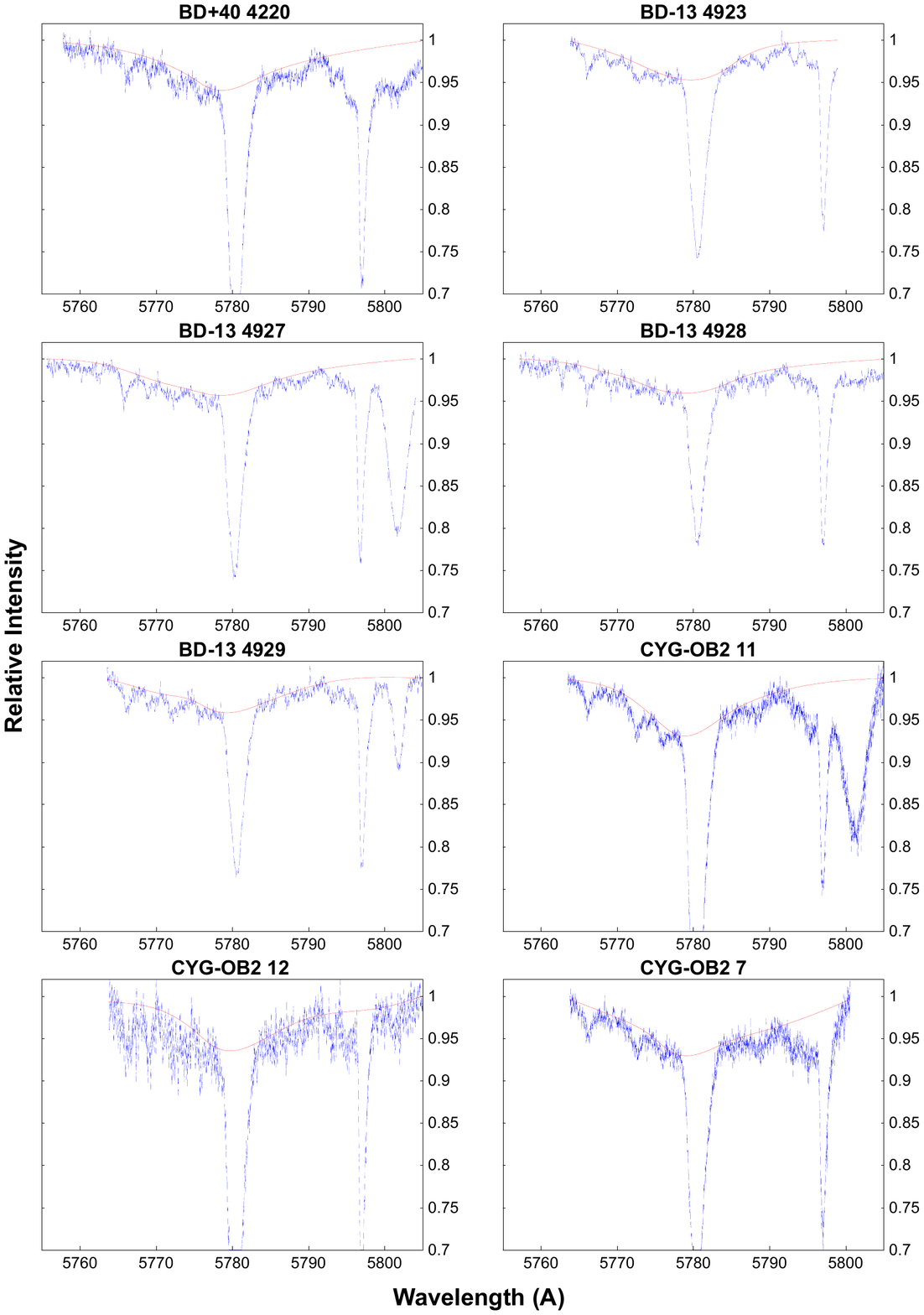}

\subsection{Profiles of 6175 DIB}

\includepdf[pages={1-5}]{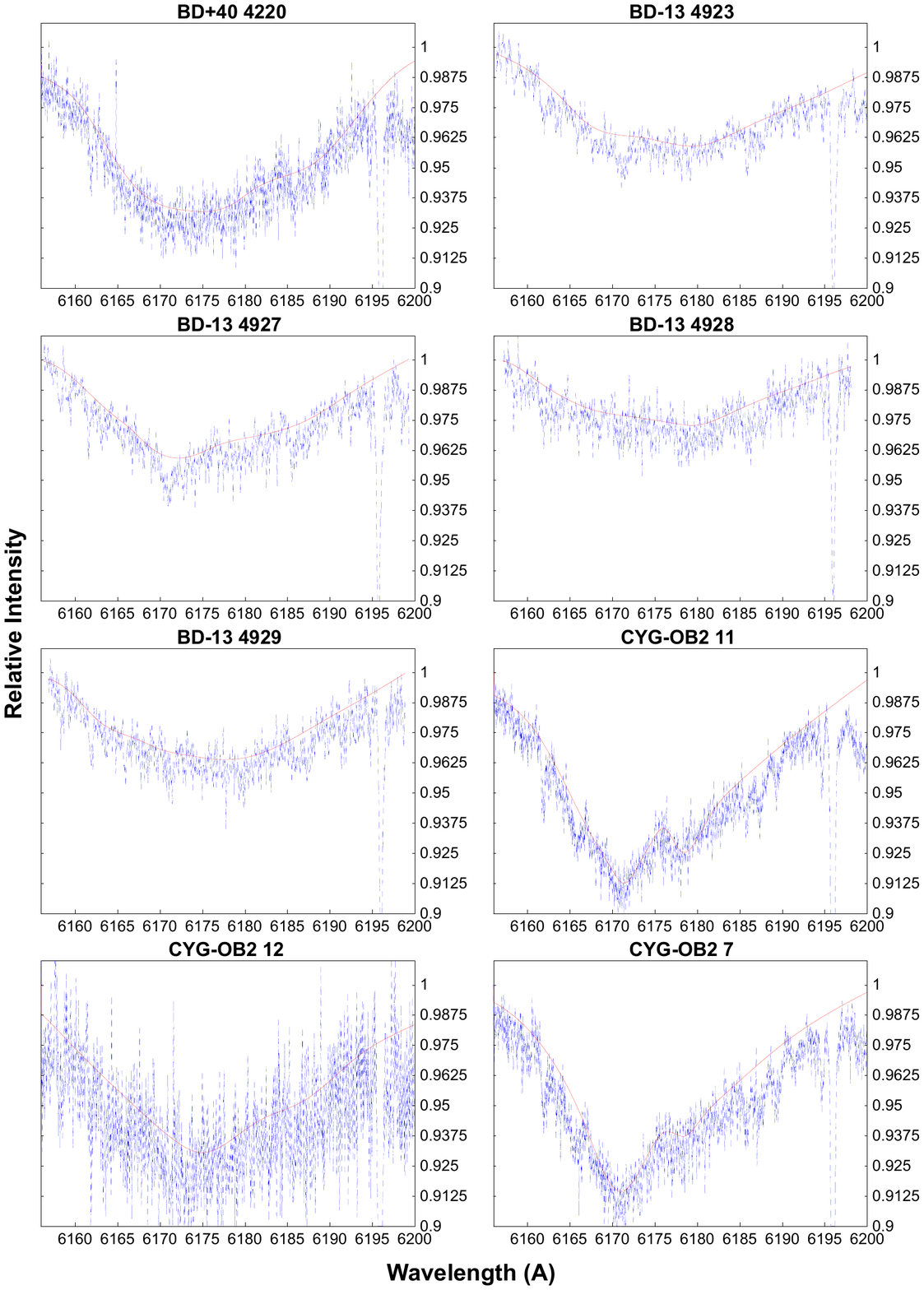}

\end{document}